# Virtual-Alloy Physics-Aware Semantic System for extrapolative prediction of perovskite solar cells

Qiwei Han[*1,2] and Chi Zhou[2]

[1]Department of Chemistry, Duke University, Durham, NC 27708, USA

[2]Department of Computer Science, Georgia Institute of Technology, Atlanta, GA 30332, USA

[*]Corresponding author.

**Abstract**

Existing machine-learning models for perovskite solar cells struggle to predict devices made from new materials. Most treat device layers as arbitrary categorical labels and represent perovskite absorbers either through featureless composition vectors or through rigid crystal graphs that require a pre-solved unit cell. As a result, they cannot generalize to unrecorded alloys or unfamiliar cell architectures. We present the Virtual-Alloy Physics-Aware Semantic System (V-PaSS), an architecture built around the physical continuity of perovskite solid solutions. V-PaSS represents arbitrary mixed-cation, mixed-halide absorbers by interpolating relaxed crystal graphs of 23 pure $ABX_3$ endpoint compounds via Vegard's law with learned bandgap bowing. For the device stack, it couples layer sequence and cell polarity (n-i-p or p-i-n) with interfacial band offsets and encodes chemical descriptions with a scientific language model. Evaluated under an institution-grouped protocol where test devices come exclusively from unseen laboratories, an untuned default configuration of V-PaSS achieves the lowest error among fourteen tested models (2.861% MAE, a 5.2% relative reduction over a gradient-boosted baseline at 3.017%, and outperforming the best tuned benchmark at 2.912%). V-PaSS pulls furthest ahead where test devices differ most from training: reducing error by 15.7–19.0% on unseen absorber alloys, by 16.1% on mixed absorbers from new laboratories, and widening its lead over the baseline by 3.6× on devices with unobserved transport layers. Combined with tree ensembles, V-PaSS sets a new state of the art across all five benchmark splits.

## 1 Introduction

Machine-learning models are increasingly tasked not merely with interpolating known materials data, but with guiding the discovery of compounds and devices that have never been fabricated [1–6]. Whether a model can extrapolate beyond its training set depends almost entirely on how its inputs are represented. A representation grounded in crystal structure and solid-state physics allows a model to interpret an unrecorded compound through what it physically shares with known materials. Categorical labels, by contrast, assign each distinct material an arbitrary index that carries no transferable chemical information.

The Open Perovskite Database [7] offers an unprecedented collection of experimental photovoltaic measurements for machine learning [8–11]. Yet training models that genuinely generalize across this archive is notoriously difficult for two reasons. The first is evaluation: because research groups publish multiple devices with identical or near-identical recipes, conventional random train–test splits suffer from pervasive data leakage, flattering models that merely memorize familiar laboratory practices. To assess true generalization, benchmarks must partition devices by experimental provenance [12]. The second challenge is physical representation. A solar cell is an integrated optoelectronic system whose efficiency is jointly dictated by the absorber, the full layer stack, and the processing recipe [12]. Furthermore, historical literature is heavily skewed toward a single benchmark composition, $MAPbI_3$ [7, 13, 14]. Without physical inductive biases, models quickly learn to overfit dominant compositions and fail when tasked with exploratory chemistries.

To resolve these challenges, we introduce the Virtual-Alloy Physics-Aware Semantic System (V-PaSS; Section 2). Instead of treating each alloy as an isolated composition or demanding an experimentally solved crystal structure, V-PaSS constructs absorbers from 23 pure $ABX_3$ endpoint crystals using the virtual crystal approximation [15, 16]. It represents each endpoint through the crystal graph of its relaxed structure, mixes their embeddings according to Vegard's law, and models non-linear bandgaps through learned bowing calibrated against photoemission measurements [17]. On the device side, V-PaSS arranges the stack by physical order and polarity (n-i-p versus p-i-n), computing conduction- and valence-band offsets across interfaces to quantify charge extraction energetics. Transport layers and processing descriptions are parsed by a scientific language model (SciBERT) [18], ensuring that chemically related materials retain related embeddings even when a specific compound never appeared in training. By coupling virtual-alloy crystallography, interface energetics, and domain language, V-PaSS treats solar cells as cohesive physical devices rather than flat tabular rows.

Across provenance-grouped benchmarks, V-PaSS demonstrates superior extrapolative accuracy. On the strictest institution-grouped split (level 3), where no test laboratory appears in training, V-PaSS reaches a power conversion efficiency (PCE) mean absolute error (MAE) of 2.861%. This represents a 5.2% relative error reduction over a gradient-boosted baseline (3.017%) and surpasses the best tuned benchmark model (2.912%) by 0.051 percentage points [12]. Crucially, V-PaSS excels precisely where the test devices diverge most from training. On absorber alloys withheld entirely from training, V-PaSS outperforms all evaluated tree-based models by 15.7–19.0%. When we ablate the virtual-alloy formulation by collapsing each alloy to its single dominant endpoint—for example, reducing $(FA_{0.80}MA_{0.15}Cs_{0.05})(Pb_{0.75}Sn_{0.25})(I_{0.80}Br_{0.20})_3$ to $FAPbI_3$—V-PaSS loses 34–43% of its lead, confirming that multi-endpoint mixing captures essential alloy physics. On the device side, V-PaSS's advantage over the gradient-boosted (GB) baseline on devices with novel transport layers is 3.6 times greater than on familiar layers. Stripping the semantic text representations causes error on novel transport materials to surge by 42%, proving that domain language embeddings supply generalizable chemical knowledge. Combining V-PaSS with tree-based models establishes a new state of the art across all five evaluation splits.

These results illustrate a broader principle for materials machine learning: when models respect physical continuity—here via Vegard solid solutions, interface energetics, and domain semantics—they can reliably extrapolate to materials and device configurations far outside their training data.

## 2 V-PaSS

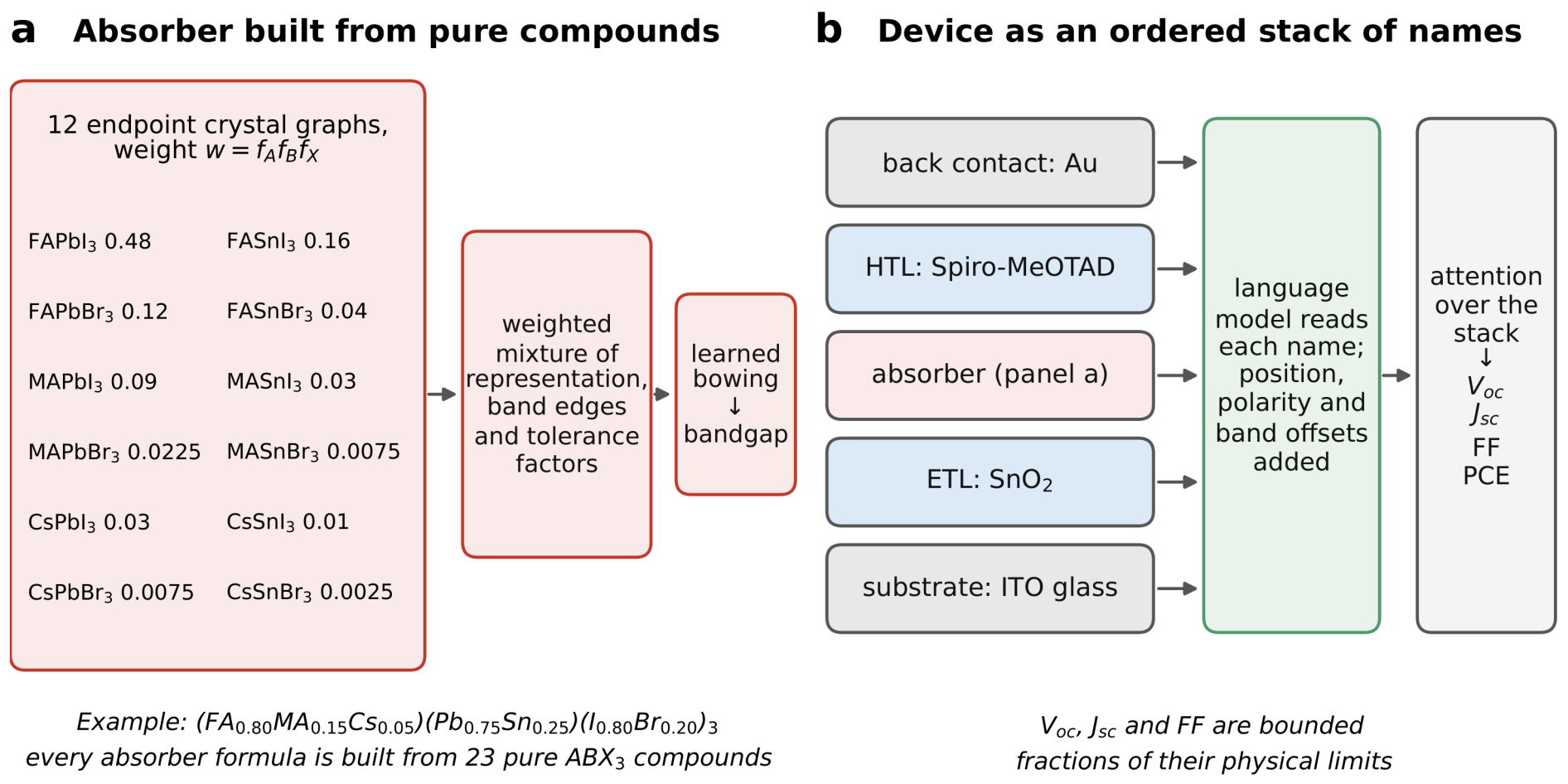

Figure 1: **V-PaSS architecture. a**, Construction of an absorber from its constituent $ABX_3$ endpoint crystals, illustrated for $(FA_{0.80}MA_{0.15}Cs_{0.05})(Pb_{0.75}Sn_{0.25})(I_{0.80}Br_{0.20})_3$: each of its twelve endpoint compounds is encoded from its relaxed crystal graph, weighted by the product of site ion fractions, and aggregated with learned bandgap bowing. **b**, Multi-layer device stack from substrate to back contact: language-model embeddings of layer descriptions, position and polarity encodings, interfacial band offsets between the absorber and transport layers, and stack cross-attention predicting $V_{oc}$, $J_{sc}$, FF, and PCE, with photovoltaic metrics constrained by physical limits.

V-PaSS couples two physical representations: a virtual-alloy graph module for the perovskite absorber and a transformer-based stack encoder for the multi-layer device (Fig. 1).

In the absorber module, any experimental stoichiometry is treated as a solid solution of 23 pure $ABX_3$ endpoint crystals (Table S1). Each endpoint $i$ receives a weight $w_i = f_A f_B f_X$ equal to the product of its site ion fractions, such that $\sum_i w_i = 1$. In the example of Fig. 1a, the mixed absorber decomposes into twelve pure endpoint crystals, with weights ranging from $w = 0.48$ for $FAPbI_3$ down to $w = 0.0025$ for $CsSnBr_3$. Each endpoint is represented by the crystal graph of its relaxed structure. Atomic states $h_v$ start from atomic number embeddings and update through graph attention layers (GATv2) [19] with residual connections and layer normalization,

$$h_v^{(l+1)} = \mathrm{ELU}\left(\mathrm{LN}\left(h_v^{(l)} + \sum_{u \in N(v)} \alpha_{vu}^{(l)} W^{(l)} h_u^{(l)}\right)\right), \tag{1}$$

where $\alpha_{vu}^{(l)}$ are attention weights across atomic neighbours $N(v)$ in layer $l$. A learned attention pooling operator aggregates atom-level states $V_i$ into an endpoint embedding,

$$p_i = W_p \sum_{v \in V_i} \mathrm{softmax}_v\left(g\left(h_v^{(L)}\right)\right) h_v^{(L)}, \tag{2}$$

where $g$ scores atomic relevance and $W_p$ projects the pooled state. Following Vegard's law, the virtual-alloy representation and linear band edges are computed as composition-weighted sums over the endpoint library,

$$z = \sum_i w_i p_i, \qquad E_c^{\mathrm{lin}} = \sum_i w_i E_c^{(i)}, \qquad E_v^{\mathrm{lin}} = \sum_i w_i E_v^{(i)}, \tag{3}$$

where endpoint band edges $E_c^{(i)}$ and $E_v^{(i)}$ are calibrated against experimental photoemission data [17]. Bandgap non-linearity is captured through learned quadratic bowing across the $A$, $B$, and $X$ sublattices,

$$E_g = E_c^{\text{lin}} - E_v^{\text{lin}} - \Omega, \qquad \Omega = 1/2 \sum_{s \in \{A,B,X\}} f_s^T B_s f_s,$$
$$E_c = E_c^{\text{lin}} - \kappa \Omega, \qquad E_v = E_v^{\text{lin}} + (1-\kappa)\Omega, \tag{4}$$

where $f_s$ is the vector of ion fractions on site $s$, $B_s$ is a symmetric bowing matrix, and $E_c$ and $E_v$ denote the resulting bowed band edges. We assign a fraction $\varkappa = 0.35$ of total bowing to the conduction band, consistent with experimental photoemission showing that composition changes in halide perovskites shift the valence-band maximum far more than the conduction-band minimum [17, 20, 21]. This virtual crystal formulation [15, 16] distinguishes V-PaSS from composition-only models that rely on isolated element embeddings [2, 3]; V-PaSS explicitly interpolates the relaxed crystal graph embeddings of whole stoichiometric compounds.

For the device stack, a frozen scientific language model (SciBERT [18]) converts the names of all layers from substrate to back contact, along with deposition conditions and additives, into dense embeddings $e$. Interfacial band offsets between the absorber and its transport layers,

$$\Delta E_C = E_c - E_{\text{LUMO}}^{\text{ETL}}, \qquad \Delta E_V = E_{\text{HOMO}}^{\text{HTL}} - E_v, \tag{5}$$

are evaluated from tabulated energy levels. These offsets are mapped to bottom and top layers according to cell polarity (n-i-p versus p-i-n), conditioning the transport layer embeddings via $e' = e + \text{MLP}([e; \Delta E])$. The full device is then structured as an ordered stack $S$ and processed by a transformer encoder into contextual representations $D$,

$$S = [e'_{\text{bot}}; c; s; e'_{\text{top}}] + P, \qquad D = \text{TransformerEncoder}(S), \tag{6}$$

where $P$ is a learned position embedding encoding layer order in the stack (for example, the n-i-p cell in Fig. 1b: ITO glass, $SnO_2$, absorber, Spiro-MeOTAD, and Au). Two absorber tokens, $c$ and $s$, describe the perovskite from complementary viewpoints:

The context token $c$ integrates chemical stoichiometry, processing descriptors, and electronic alignment,

$$q = \text{MLP}([e_f; e_{\text{dep}}; \varphi]), \qquad c = \text{LN}(q + \text{Attn}(q, \{p_i\}_{w_i>0})) + \sigma(W_g \acute{e}_{\text{add}}) \odot e_{\text{add}},$$
$$\varphi = [E_c, E_v, E_g, E_{\text{LUMO}}^{\text{ETL}}, E_{\text{HOMO}}^{\text{HTL}}, \Delta E_C, \Delta E_V, \Omega, S_X, \delta_A, \tau_A, a_{\text{pin}}, a_{\text{meas}}, E_g^{\text{rec}} - E_g],$$
$$S_X = -\sum_x f_x \ln f_x, \qquad \delta_A = \left[\sum_a f_a (1 - r_a/\acute{r}_A)^2\right]^{1/2}, \qquad \acute{r}_A = \sum_a f_a r_a, \tag{7}$$

where $e_f$ and $e_{\text{dep}}$ are language-model embeddings of the chemical formula and deposition description. The descriptor vector $\varphi$ gathers bowed band parameters, interfacial offsets, halide mixing entropy $S_X$, $A$-site size variance $\delta_A$, learned cation factor $\tau_A$, binary flags for device polarity and bandgap measurement, and the discrepancy with recorded bandgap $E_g^{\text{rec}}$. The term Attn executes cross-attention from query $q$ over active endpoint embeddings, while additive embeddings $e_{\text{add}}$ are gated through a learned sigmoid projection.

The structural instability token $s$ penalizes endpoints that depart from stable perovskite geometry, isolating structural components orthogonal to the mean alloy vector $z$,

$$u_i = 1 - \exp\left[-\left(\frac{t_i - t_0}{\Delta t}\right)^2\right], \qquad \tilde{s} = \sum_i w_i u_i p_i, \qquad s = \tilde{s} - (\tilde{s} \cdot \hat{z})\hat{z}, \tag{8}$$

where $\hat{z}=z/\|z\|$ and $t_i=(r_A+r_X)/[\sqrt{2}(r_B+r_X)]$ is the Goldschmidt tolerance factor of endpoint $i$ [22]. Calculated from Shannon and effective ionic radii [23, 24], $t_i$ departs from optimal packing parameters $t_0$ = 0.93 and $\Delta t$ = 0.05 as lattice distortion increases; the instability score $u_i$ rises accordingly, whereas $s$ vanishes for pure, unmixed compounds.

In the output heads, each metric $k \in \{V, J, F, P\}$ is scored by attending from the encoded stack back to the endpoint embeddings,

$$H_k=\mathrm{LN}\left(D+\mathrm{Attn}_k(D,\{w_i p_i\})\right), \qquad H_k^{'}=\mathrm{LN}\left(H_k+\mathrm{Attn}_k^{'}(H_k,\{w_i u_i p_i\})\right),$$
$$y_k=\mathrm{MLP}_k\left([z;\mathrm{vec}(H_k^{'});\tilde{x}_{\mathrm{tab}};\tilde{e}_{\mathrm{sub}};\tilde{e}_{\mathrm{bc}}]\right), \tag{9}$$

where $\tilde{x}_{\mathrm{tab}}$, $\tilde{e}_{\mathrm{sub}}$, and $\tilde{e}_{\mathrm{bc}}$ project tabular fabrication descriptors, substrate embeddings, and back-contact embeddings. To guarantee that predictions obey basic physical bounds, output scores are mapped to thermodynamic limits under detailed balance [25, 26],

$$V_{oc}=\sigma(y_V)(E_g^{\mathrm{rec}}-\delta_V), \qquad J_{sc}=\sigma(y_J)J(E_g^{\mathrm{rec}}), \qquad FF=FF_{max}\sigma(y_F),$$
$$\widetilde{\mathrm{PCE}}=y_P, \tag{10}$$

where $\sigma$ is the sigmoid activation, $J(E_g)$ represents detailed-balance photocurrent under AM1.5G illumination, $\delta_V$ = 0.25 V accounts for fundamental radiative voltage loss, and $FF_{max}$=0.90 sets the thermodynamic fill factor upper bound [27–29]. Multi-task training optimizes a Smooth L1 loss across all heads, weighting sub-metric losses and an efficiency consistency term through learned homoscedastic uncertainty [30],

$$L=\ell\left(y_P,\widetilde{\mathrm{PCE}}^m\right)+\sum_{k\in\{V,J,F\}}\left[e^{-\gamma_k}\ell(\sigma(y_k),\rho_k)+\gamma_k\right]+e^{-\gamma_c}\ell\left(\widetilde{V_{oc}J_{sc}FF},\widetilde{\mathrm{PCE}}^m\right)+\gamma_c$$
$$+\lambda_g\ell_m\left(E_g,E_g^{\mathrm{rec}}\right)+\lambda_e\sum_i\ell\left(\pi(p_i),\hat{\pi}_i\right), \tag{11}$$

where $\ell$ denotes Smooth L1 loss, $\rho_k$ denotes measured metrics scaled to theoretical bounds, and $\gamma$ represents homoscedastic uncertainty weights. Auxiliary supervision terms $\ell_m$ and $\lambda_e$ ensure that learned bandgaps and endpoint embeddings align with measured band edges and structural factors (Methods).

# 3 Results and discussion

## 3.1 Performance under provenance-grouped evaluation

We benchmarked V-PaSS against thirteen comparison models across the five-split provenance protocol of ref. [12]: a random split, three hierarchical provenance splits (levels 1–3), and a temporal split. Level 3 (the institution-grouped split) provides the definitive test of laboratory-level extrapolation, ensuring that train and test sets share no research institutions.

Table 1 compares test error across models on identical test partitions, with error reported as mean absolute error (MAE) in percentage points of PCE. Fig. 2a ranks all fourteen models at level 3. Comprehensive performance across splits and paired statistical differences are provided in Tables S2 and S3.

Table 1: **Model performance across provenance-grouped evaluation splits.** PCE mean absolute error (%) on identical test sets across splits. Benchmark models and error values are from ref. [12]; the gradient-boosted baseline uses fixed hyperparameters. V-PaSS and the baseline average five training seeds on the random split and three seeds on grouped and temporal splits. The final column shows V-PaSS combined with six tuned tree ensembles (A4), weighted by validation least absolute error (Methods; Table S4; on the temporal split, pairing V-PaSS with XGBoost trained on V-PaSS's physical features, denoted A3, achieves the lowest overall error at 3.189%). Paired differences with bootstrap standard errors are in Table S3. Bold indicates lowest error per row.

| **Split** | **Baseline** | **V-PaSS** | **Best benchmark model (MAE)** | **V-PaSS + trees** |
|---|---|---|---|---|
| Level 3 | 3.017 | 2.861 | SciBERT+attention (2.912) | **2.833** |
| Level 2 | 2.869 | 2.876 | XGBoost (2.832) | **2.775** |
| Level 1 | 2.457 | 2.502 | XGBoost (2.396) | **2.375** |
| Random | 2.131 | 2.040 | LightGBM (1.973) | **1.939** |
| Temporal | 3.462 | 3.236 | MLP (**3.192**) | 3.248 |

Because all models were evaluated on identical test devices, we analyzed paired error differences with bootstrap standard errors clustered by test institution (Methods; Table S3). V-PaSS leads the field across all thirteen benchmark models at level 3: it achieves a 5.2% relative error reduction over the gradient-boosted baseline (an MAE reduction of −0.156 ± 0.057 percentage points, corresponding to 2.7 standard errors), outperforms the tuned tree ensembles by 2.5–5.3% relative (0.073–0.161 percentage points in MAE), and edges out the strongest neural benchmark (SciBERT with attention) by 0.051 ± 0.051 percentage points (Table S3).

This comparison highlights a fundamental split in model behavior. On random splits and at levels 1 and 2, where test devices share duplicates, publications, or institutions with the training set, tree ensembles exploit memorization. But at level 3, where provenance leakage is completely closed, V-PaSS's physical and semantic inductive biases win out. Strikingly, V-PaSS achieved this performance with an untuned default architecture, whereas each of the twelve benchmark models was selected from an extensive hyperparameter search across 30–60 configurations [12].

Combining models confirms that physical inductive biases and empirical tree recall offer complementary predictive information. Stacking V-PaSS with the six tuned tree models using non-negative validation weights lowers test error across all splits (Tables 1 and S4), establishing a new state of the art at levels 1–3 and on the random split. On the temporal split, combining V-PaSS with XGBoost trained on V-PaSS's physical input representations (combination A3) yields the lowest overall error across all models (3.189% MAE, Table S4), outperforming the best single benchmark model (MLP, 3.192%).

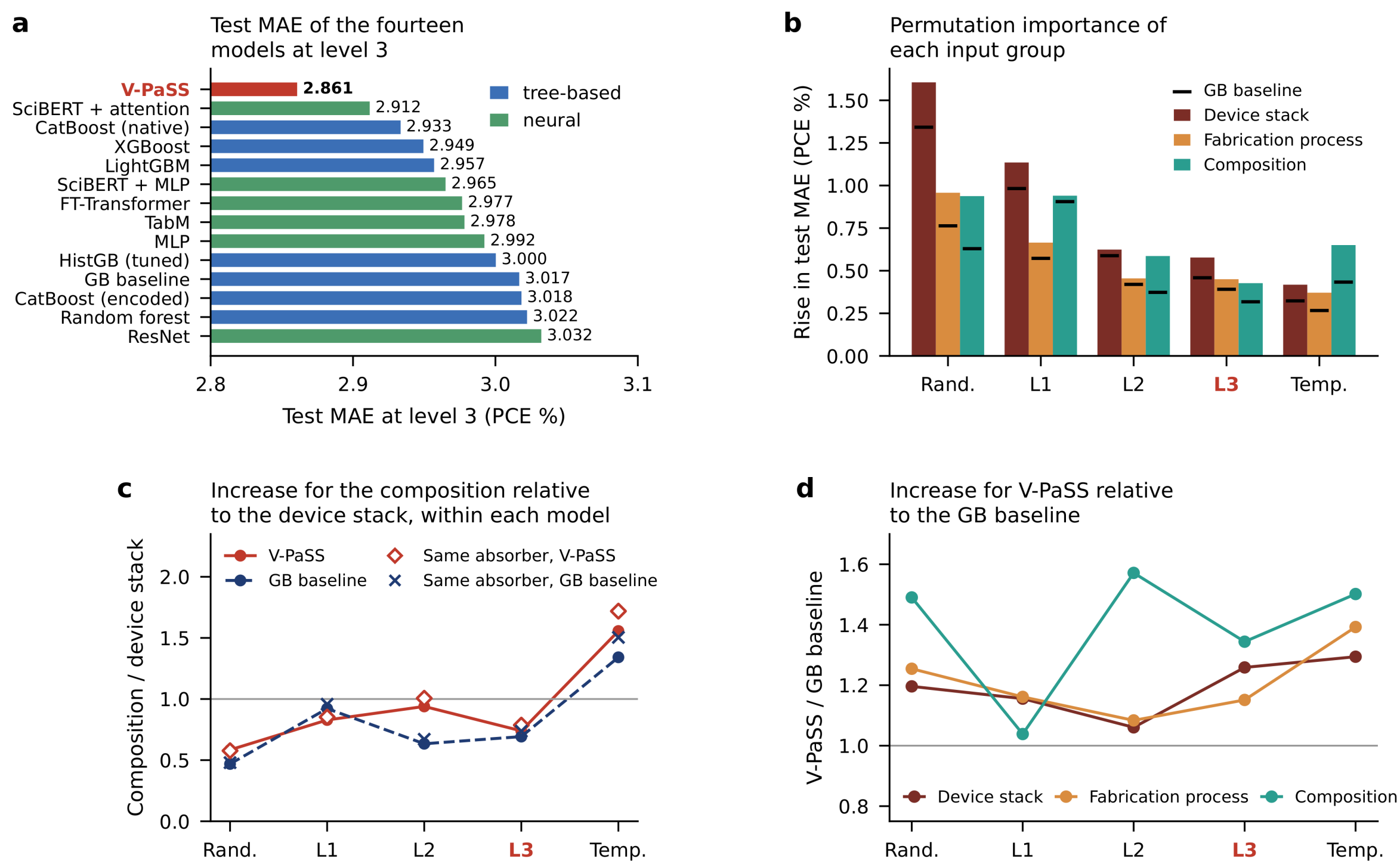

Figure 2: **Provenance-grouped benchmark performance and feature modality reliance. a**, Test MAE of fourteen models at level 3. **b**, Permutation importance of input modalities (PCE MAE increase upon shuffling device stack, fabrication process, or composition), evaluated on affected test devices for V-PaSS (bars) and the GB baseline (ticks). **c**, Ratio of composition importance to stack importance for V-PaSS (solid) and baseline (dashed); symbols mark ratios evaluated solely within identical absorbers. **d**, Ratio of V-PaSS importance to baseline importance across input modalities. Data: Tables 1, S2, and S5.

## 3.2 Input modality importance across evaluation splits

To understand what drives V-PaSS's accuracy, we evaluated permutation feature importance for both V-PaSS and the gradient-boosted baseline [31]. We permuted the composition, device stack, and fabrication process independently (Fig. 2b–d; Table S5; Methods). Because $MAPbI_3$ devices dominate the archive, randomly shuffling compositions leaves roughly half of the test assignments unchanged; we therefore scored importance specifically on the devices whose inputs actually changed under permutation, preventing artificial dilution of the importance scores.

For both models, the device stack is the most influential input group on the random split and at levels 1–3, whereas the composition becomes dominant on the temporal split (Fig. 2b). Tracking the ratio of composition importance to stack importance shows that composition becomes steadily more critical as the evaluation split pushes further from the training data (Fig. 2c and Fig. S1). Shuffling the device stack exclusively among devices with the same absorber leaves this trend intact (symbols in Fig. 2c; Table S5), ruling out the possibility that the shift is an artifact of composition–stack correlation.

This shift directly reflects how experimental research evolved over time (Table S6). Between level 1 and the temporal split, the fraction of test devices with unobserved absorber compositions multiplies nearly ninefold, while the fraction with novel stacks only doubles. As perovskite research pushed into multinary alloys, predicting device performance increasingly required interpreting compositions that had never been seen before. V-PaSS handles this by constructing unobserved alloys from known crystal endpoints, whereas label-based models must guess from target-encoded historical averages.

Comparing modality importances between models demonstrates that V-PaSS derives a higher proportion of its accuracy from physical and chemical features than does the baseline (Fig. 2d). For all three modalities, the ratio of V-PaSS importance to baseline importance exceeds 1 across every split. The gradient-boosted baseline substitutes layers and absorbers with historical target-encoded means, relying heavily on empirical recall. By contrast, V-PaSS models novel compositions by projecting them onto constituent crystal endpoints. Consequently, the reliance of V-PaSS on composition features scales directly with the incidence of unobserved alloys in the test set.

### 3.3 Extrapolation to novel device architectures and transport layers

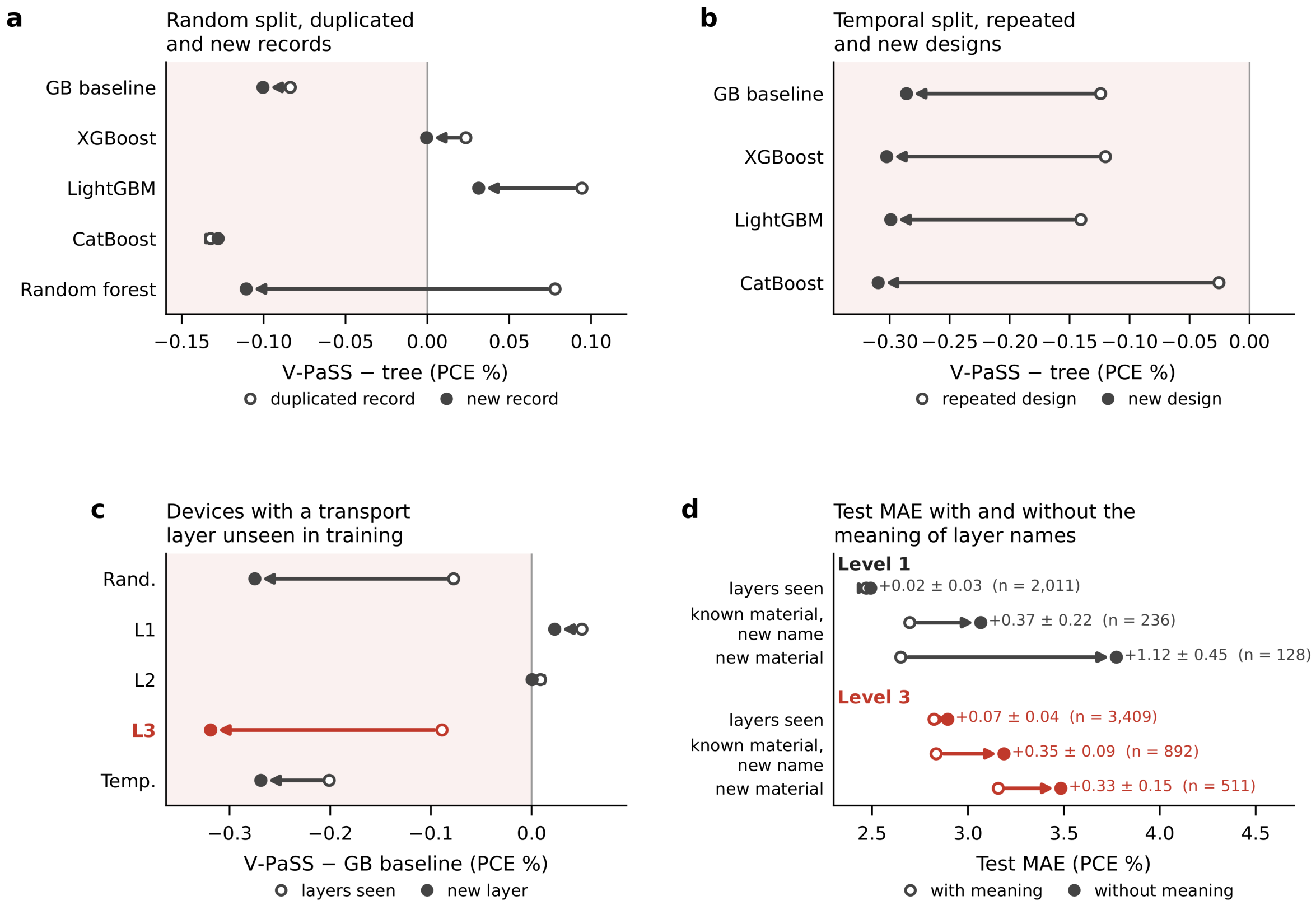

Figure 3: **Generalization on unobserved device architectures and transport materials.** In **a**–**c**, arrows track paired MAE differences from familiar (open circle) to novel (filled circle) test devices; leftward vectors indicate growing V-PaSS advantage. **a**, Random split: exact training duplicates versus unique devices across tree-based models (Table S7). **b**, Temporal split: pre-2016 architecture repeats versus novel device stacks (Table S8). **c**, Devices with familiar versus novel transport layers (ETL or HTL unseen in training), compared to the GB baseline across splits (Table S9). **d**, Test MAE of V-PaSS with language-model semantics (open circle) and with every name replaced by a fixed random token that keeps the layer's identity (filled circle) at levels 1 and 3, for devices with familiar transport layers, with an unseen layer whose materials are all known (a known material under a new name), and with an unseen layer containing a new material; labels give the MAE increase ± one standard error and the number of devices (Table S11).

Next, we examined where V-PaSS's architectural advantage on the device stack comes from. On the random split, LightGBM and XGBoost outperform V-PaSS overall. But partitioning the test records into exact duplicates (56.9%) and unique devices reveals that their lead is entirely an artifact of database repetition (Fig. 3a; Table S7). Once evaluated on unique devices, the tree ensembles' recall advantage collapses: V-PaSS matches XGBoost (2.244% MAE, difference of −0.000 ± 0.044 percentage points), overtakes Random Forest by 0.111 ± 0.048 percentage points, leads CatBoost and the baseline by 0.100–0.128 percentage points (4.3–5.4% relative reduction), and cuts LightGBM's lead by two thirds (to 0.031 ± 0.044 percentage points).

A similar dynamic appears on the temporal split, where V-PaSS achieves 4–11% lower error than every tuned tree ensemble (Table S2). When we separate post-2018 devices into novel architectures and those repeating pre-2016 stacks (37.1%; Fig. 3b; Table S8), V-PaSS's margin over the baseline, LightGBM, and XGBoost more than doubles on novel designs, reaching an MAE lead of 0.286–0.303 percentage points (8.0–8.4% relative error reduction) and widening its lead over CatBoost twelvefold (to 0.310 ± 0.063 percentage points; Table S8a). Tree ensembles compete effectively when a device repeats an existing recipe, but fall 0.261–0.267 percentage points behind when required to extrapolate to new designs (Table S8b).

We then tested whether language-model embeddings allow V-PaSS to generalize to unobserved transport layers (Fig. 3c; Table S9). On the 1,403 devices at level 3 with an unseen ETL or HTL, V-PaSS's lead over the baseline expands to 3.6 times its margin on familiar devices: an MAE lead of 0.319 ± 0.109 percentage points (a 10% relative error reduction) versus 0.089 ± 0.052 percentage points on familiar layers. At levels 1 and 2, this difference is masked across the full test set because sibling devices in training provide proxy information (79.1% share a publication at level 1, and 89.7% an institution at level 2; Table S10). But on the isolated subset of devices without related training records, V-PaSS consistently outperforms the baseline by 0.250–0.295 percentage points across both levels (Table S10).

To confirm that this capability relies on chemical language semantics rather than simple token indexing, we retrained V-PaSS with text fields replaced by fixed random token sequences, preserving categorical identities and numerical features while stripping semantic meaning (Methods; Fig. 3d; Table S11). At level 3, removing the meaning of layer names barely alters error on devices whose transport layers were seen in training (+0.07 ± 0.04 percentage points), because a familiar layer is recognized by its identity alone. The cost falls almost entirely on devices with an unseen layer: approximately +0.35 percentage points, five times the familiar-device rise. This penalty is virtually identical whether the unseen name represents a known material under a new designation (+0.35 ± 0.09 percentage points) or a material absent from training altogether (+0.33 ± 0.15 percentage points), showing that the language model supplies the chemical meaning of any unfamiliar name. Across all unseen-layer devices at level 3, this error increase (+0.344 ± 0.078 percentage points) accounts for V-PaSS's entire lead over the gradient-boosted baseline on those devices (0.319 ± 0.109 percentage points). At level 1, the increase for new materials appears larger (+1.12 ± 0.45 percentage points) because most of these devices share a publication with training devices, which the model can relate to a new layer only through the meaning of its name; at level 3, where no such shared-publication devices exist, the increase is the same for new materials and for known materials under new names (Fig. 3d). Domain language representations are therefore the direct driver of V-PaSS's generalization across unobserved transport materials.

## 3.4 Extrapolation to unseen absorber alloys

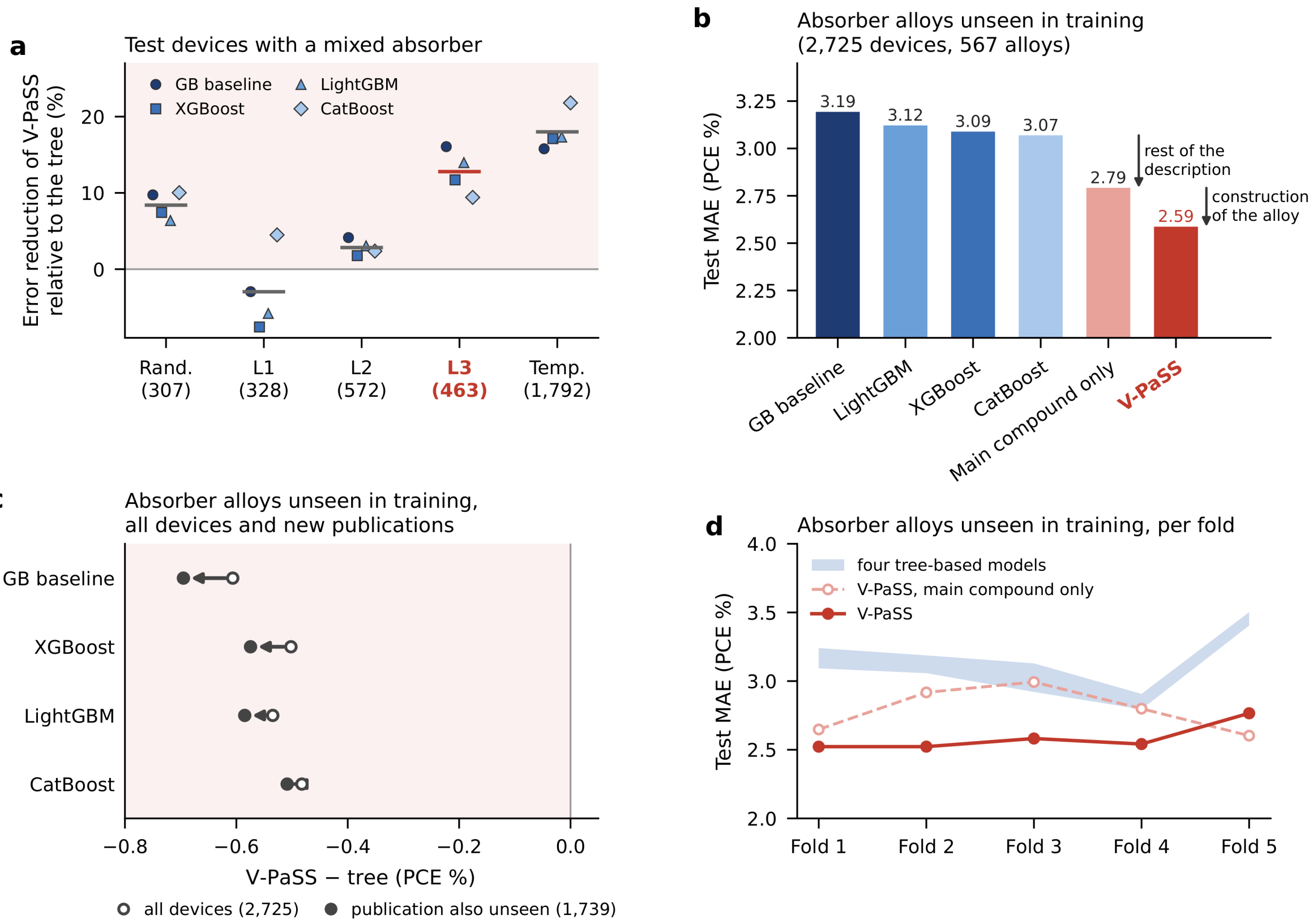

Figure 4: **Extrapolative performance on mixed absorbers and novel absorber alloys. a**, Percentage error reduction of V-PaSS relative to tree-based models on mixed-absorber devices across splits. **b–d**, Generalization on unobserved absorber alloys (2,725 devices across 567 alloys, 5-fold cross-validation without measured bandgap). **b**, Test MAE across tree ensembles, V-PaSS with alloys collapsed to dominant endpoints, and full V-PaSS; arrows mark performance gains from full device encoding versus multi-endpoint alloy mixing. **c**, MAE advantage over tree models on all alloy devices versus devices whose publication is also unobserved. **d**, Test MAE across the five cross-validation folds. Data: Tables S12 and S13.

The virtual-alloy formulation specifically targets mixed-perovskite absorbers. On the 463 mixed-absorber devices at level 3, V-PaSS reduces error by 16.1% relative to the baseline (an MAE lead of 0.470 ± 0.184 percentage points) and outperforms the three tuned tree models by 9.4–14.0% (0.255–0.399 percentage points in MAE; Fig. 4a; Table S12). On the temporal split, this advantage holds against the baseline (15.8%, lead of 0.575 ± 0.112 percentage points) and widens against the tuned tree models to 17.1–21.8% (0.633–0.855 percentage points in MAE; Table S12).

To test alloy extrapolation under the strictest possible conditions, we held out entire alloy compositions using a 5-fold cross-validation scheme across all 2,725 mixed-absorber devices in the database, ensuring that none of the 567 distinct alloys appeared in both training and test sets (Methods; Table S13). All pure-composition devices were retained in training as elemental building blocks, and measured bandgaps were withheld from all models. V-PaSS reaches an MAE of 2.587%, outperforming the four tree models (3.069–3.193%) with a 15.7–19.0% lower error (Fig. 4b). V-PaSS is more accurate than every comparator model in all five folds (Fig. 4d), and its margin over tree models widens from 0.48–0.61 to 0.51–0.70 percentage points on the 1,739 devices whose publications are also absent from training (Fig. 4c; Table S13).

To isolate how much of this advantage comes from virtual-alloy construction versus the rest of the architecture, we retrained V-PaSS on the same folds with each alloy collapsed to its dominant endpoint (Methods; Fig. 4b,d; Table S13). Even when given only the primary endpoint compound, V-PaSS still outperforms all four tree ensembles by 9–13%, demonstrating the strength of its device stack and interface modeling. Reconstructing the full alloy via multi-endpoint Vegard mixing lowers error by a further 7% across all 2,725 devices (and by 0.320 ± 0.080 percentage points in MAE, an 11.4% relative reduction, excluding the single outlier alloy $CsPbBr_2I$, 61% of whose 92 devices stem from a single early study reporting 6% efficiency, well below what the composition physically implies; Table S14). Even on this outlier alloy, V-PaSS decisively outperforms every tree ensemble (5.15% MAE versus 7.40–7.80%). Examining $CsPbBr_2I$ also exposed a transcription error in the source database. One of its 92 devices is in fact a zinc-substituted $CsPbI_2Br$ cell whose halide stoichiometry was inverted upon database entry [32]. Its recorded bandgap of 1.88 eV lies 0.21 eV below the 2.09 eV that V-PaSS constructs for $CsPbBr_2I$, and its recorded photocurrent (15.8 $\mathrm{mA\,cm^{-2}}$) exceeds the theoretical detailed-balance limit at that bandgap. Because V-PaSS grounds each stoichiometry in solid-state band physics rather than categorical labels, it intrinsically establishes physically expected optoelectronic limits for each composition, enabling the identification of records whose reported measurements physically contradict their assigned formulas; deploying this capability for systematic archive auditing is left to future work. Multi-endpoint alloy construction accounts for 34–43% of V-PaSS's overall margin over tree models, proving that continuous solid-solution modeling is indispensable for compositional extrapolation.

# 4 Conclusion

In summary, we introduced V-PaSS to tackle the problem of extrapolative prediction in perovskite photovoltaics. By constructing absorbers from pure $ABX_3$ endpoint crystals through Vegard's law with learned bowing, and by encoding device stacks through layer position, polarity, interfacial band offsets, and scientific language models, V-PaSS replaces tabular memorization with physical continuity.

Across rigorous provenance-grouped benchmarks, an untuned default configuration of V-PaSS outperforms thirteen competitive models, reducing error by 5.2% relative to the baseline on the strictest institution-grouped split. Its performance advantage grows precisely where experimental discovery needs it most: on novel cell architectures and mixed absorbers, where it cuts error by 8–22% relative to tuned tree ensembles; on unobserved transport layers, where chemical language embeddings avert a 42% error surge; and on unrecorded absorber alloys, where virtual-alloy delivers a 15.7–19.0% error reduction across all cross-validation folds. Ensembling V-PaSS with tree-based models establishes a new state of the art across all five benchmark splits.

More broadly, this work demonstrates that the extrapolative horizon of materials machine learning is governed by representation design. Grounding models in physical solid-solution theory and domain language offers a practical roadmap for accelerating the discovery of materials and devices well beyond existing experimental archives.

**Acknowledgements.** The authors used the large language models Claude (Anthropic) and Gemini (Google) as assistants in drafting parts of the manuscript and the analysis code. All text, code and results were reviewed, verified and revised by the authors, who take full responsibility for the content of this article.

**Competing interests.** The authors declare no competing interests.

## Supporting Information

Table S1: Endpoint band edges and their sources. VBM and CBM are the valence-band maximum and the conduction-band minimum on the vacuum scale; V-PaSS computes the gap of every absorber from them (CBM minus VBM). The optical gap is the value reported in the same source, listed for reference; it is not an input. [a]Valence band as measured; conduction band set so that CBM minus VBM is 1.36 eV, the midpoint of the reported optical range of 1.31–1.41 eV, since the inverse-photoemission edge is the less precise of the two. [b]Assigned PBE-DFT band edges from ref. [6] with a single global vacuum shift during model training (Methods); no experimental photoemission measurement is available (together <0.01% of archive composition mass).

| Endpoint | VBM (eV) | CBM (eV) | Optical gap (eV) | Source |
|---|---|---|---|---|
| $MAPbI_3$ | −5.93 | −4.36 | 1.59 | Tao et al. [17] |
| $FAPbI_3$ | −6.24 | −4.74 | 1.51 | Tao et al. [17] |
| $CsPbI_3$ | −6.25 | −4.47 | 1.72 | Tao et al. [17] |
| $MAPbBr_3$ | −6.60 | −4.25 | 2.30 | Tao et al. [17] |
| $FAPbBr_3$ | −6.70 | −4.51 | 2.25 | Tao et al. [17] |
| $CsPbBr_3$ | −6.53 | −4.17 | 2.31 | Tao et al. [17] |
| $CsPbCl_3$ | −6.80 | −3.77 | 2.99 | Tao et al. [17] |
| $FAPbCl_3$ | −6.94 | −3.98 | 3.02 | Tao et al. [17] |
| $MASnI_3$ | −5.39 | −4.07 | 1.24 | Tao et al. [17] |
| $FASnI_3$ | −5.34 | −3.98 | 1.36 | Tao et al. [17][a] |
| $CsSnI_3$ | −5.69 | −4.38 | 1.25 | Tao et al. [17] |
| $MASnBr_3$ | −5.67 | −3.42 | 2.13 | Tao et al. [17] |
| $FASnBr_3$ | −6.23 | −3.60 | 2.63 | Tao et al. [17] |
| $CsSnBr_3$ | −5.82 | −4.07 | 1.81 | Tao et al. [17] |
| $MASnCl_3$ | −6.85 | −3.36 | 3.50 | Tao et al. [17] |
| $RbPbI_3$ | −6.25 | −4.47 | 1.72 | Tao et al. [17], as $CsPbI_3$ |
| $KPbI_3$ | −6.25 | −4.47 | 1.72 | Tao et al. [17], as $CsPbI_3$ |
| $RbPbBr_3$ | −6.53 | −4.17 | 2.31 | Tao et al. [17], as $CsPbBr_3$ |
| $CsGeI_3$ | −5.69 | −4.38 | 1.25 | Tao et al. [17], as $CsSnI_3$ |
| $MAGeI_3$ | −5.39 | −4.07 | 1.24 | Tao et al. [17], as $MASnI_3$ |
| $CsGeBr_3$ | −5.82 | −4.07 | 1.81 | Tao et al. [17], as $CsSnBr_3$ |
| $MACuI_3$ | —[b] | —[b] | —[b] | Materials Project [6], shifted |
| $CsBaI_3$ | —[b] | —[b] | —[b] | Materials Project [6], shifted |

Table S2: Test MAE (%) of V-PaSS and the thirteen benchmark models on every split, ordered by error at level 3. The errors of the benchmark models are those of the previous study [12]. Lowest error on each split in bold.

| Model | Random | Level 1 | Level 2 | Level 3 | Temporal |
|---|---|---|---|---|---|
| V-PaSS | 2.040 | 2.502 | 2.876 | **2.861** | 3.236 |
| SciBERT+attention | 1.993 | 2.454 | 2.871 | 2.912 | 3.228 |
| CatBoost (native) | 2.171 | 2.536 | 2.839 | 2.933 | 3.440 |
| XGBoost | 2.027 | **2.396** | **2.832** | 2.949 | 3.471 |
| LightGBM | **1.973** | 2.409 | 2.848 | 2.957 | 3.476 |
| SciBERT+MLP | 2.080 | 2.563 | 2.901 | 2.965 | 3.259 |
| FT-Transformer | 2.078 | 2.701 | 2.962 | 2.977 | 3.215 |
| TabM | 2.041 | 2.545 | 2.965 | 2.978 | 3.246 |
| MLP | 2.065 | 2.683 | 2.938 | 2.992 | **3.192** |
| Histogram gradient boosting | 2.025 | 2.429 | 2.863 | 3.000 | 3.521 |
| Gradient-boosted baseline | 2.131 | 2.457 | 2.869 | 3.017 | 3.462 |
| CatBoost (encoded) | 2.127 | 2.480 | 2.892 | 3.018 | 3.388 |
| Random forest | 2.044 | 2.481 | 2.923 | 3.022 | 3.621 |

| Model | Random | Level 1 | Level 2 | Level 3 | Temporal |
|---|---|---|---|---|---|
| ResNet-MLP | 2.084 | 2.718 | 3.033 | 3.032 | 3.268 |

Table S3: V-PaSS minus each benchmark model (test MAE, %; negative means V-PaSS is more accurate), ± one standard error over the test units of each split: 2,095 groups on the random split, 1,258 groups at level 1, 873 publications at level 2, 169 institutions at level 3 and 2,071 publications on the temporal split.

| V-PaSS minus | Random | Level 1 | Level 2 | Level 3 | Temporal |
|---|---|---|---|---|---|
| SciBERT+attention | +0.048 ± 0.025 | +0.049 ± 0.049 | +0.004 ± 0.044 | −0.051 ± 0.051 | +0.007 ± 0.041 |
| CatBoost (native) | −0.130 ± 0.027 | −0.034 ± 0.049 | +0.037 ± 0.050 | −0.073 ± 0.041 | −0.204 ± 0.045 |
| XGBoost | +0.013 ± 0.024 | +0.107 ± 0.041 | +0.044 ± 0.048 | −0.089 ± 0.050 | −0.235 ± 0.044 |
| LightGBM | +0.067 ± 0.025 | +0.094 ± 0.041 | +0.028 ± 0.049 | −0.096 ± 0.050 | −0.240 ± 0.045 |
| SciBERT+MLP | −0.040 ± 0.023 | −0.061 ± 0.049 | −0.025 ± 0.048 | −0.104 ± 0.052 | −0.024 ± 0.037 |
| FT-Transformer | −0.038 ± 0.029 | −0.198 ± 0.073 | −0.086 ± 0.051 | −0.116 ± 0.045 | +0.021 ± 0.045 |
| TabM | −0.000 ± 0.027 | −0.043 ± 0.072 | −0.089 ± 0.051 | −0.117 ± 0.045 | −0.010 ± 0.044 |
| MLP | −0.024 ± 0.030 | −0.180 ± 0.088 | −0.062 ± 0.046 | −0.131 ± 0.051 | +0.044 ± 0.045 |
| Histogram gradient boosting | +0.015 ± 0.024 | +0.074 ± 0.043 | +0.012 ± 0.052 | −0.140 ± 0.060 | −0.286 ± 0.045 |
| Gradient-boosted baseline | −0.091 ± 0.025 | +0.046 ± 0.044 | +0.006 ± 0.051 | −0.156 ± 0.057 | −0.226 ± 0.044 |
| CatBoost (encoded) | −0.087 ± 0.025 | +0.022 ± 0.046 | −0.016 ± 0.048 | −0.158 ± 0.047 | −0.152 ± 0.046 |
| Random forest | −0.003 ± 0.027 | +0.021 ± 0.046 | −0.048 ± 0.053 | −0.161 ± 0.059 | −0.385 ± 0.050 |
| ResNet-MLP | −0.044 ± 0.032 | −0.216 ± 0.065 | −0.157 ± 0.056 | −0.171 ± 0.055 | −0.033 ± 0.045 |

Table S4: Test MAE (%) of V-PaSS combined with tree-based models. A1: with the gradient-boosted (GB) baseline; A2: with XGBoost; A3: with XGBoost on V-PaSS's input representation; A4: non-negative combination of V-PaSS and the six tuned tree-based models. A4, which has a lower error than the best benchmark model on the random split and at levels 1 to 3, is the one reported in the main text, while A3 achieves the lowest error on the temporal split (3.189%). The best single model is the most accurate of V-PaSS, the benchmark models and XGBoost on V-PaSS's input representation, which is the best single model at level 2; the best benchmark model is that of the previous study [12]. The last row gives A4 minus the best single model on each split. Lowest error on each split in bold.

| | Random | Level 1 | Level 2 | Level 3 | Temporal |
|---|---|---|---|---|---|
| A1 | 2.007 | 2.407 | 2.790 | 2.847 | 3.225 |
| A2 | 1.974 | **2.372** | 2.774 | 2.838 | 3.237 |
| A3 | 1.980 | 2.383 | **2.763** | 2.845 | **3.189** |
| A4 | **1.939** | 2.375 | 2.775 | **2.833** | 3.248 |
| best single model | 1.973 | 2.396 | 2.824 | 2.861 | 3.192 |
| best benchmark model | 1.973 | 2.396 | 2.832 | 2.912 | 3.192 |
| A4 − best single | −0.034 ± 0.010 | −0.021 ± 0.023 | −0.049 ± 0.025 | −0.027 ± 0.014 | +0.056 ± 0.037 |

Table S5: Permutation importance, the rise in test MAE (%) when the composition, the device stack or the fabrication process is permuted among the test devices, one group at a time, on every split; the ratio of the importance for V-PaSS to that for the GB baseline; and the ratios between the importances of two groups within each model. Each importance is scored on the test devices that the permutation gives a different input, which corrects for the many test devices that share the same absorber. In the rows marked same absorber, the device stack is permuted only among test devices with the same absorber composition. Each model averages its published seeds (V-PaSS: five on the random split and at level 1, three at levels 2 and 3 and on the temporal split; GB baseline: five on the random split, three on the others), with five permutations per seed. Seed-to-seed SD is at most 0.088%.

| Inputs permuted | Random | Level 1 | Level 2 | Level 3 | Temporal |
|---|---|---|---|---|---|
| *V-PaSS* | | | | | |
| Composition | 0.938 | 0.940 | 0.586 | 0.427 | 0.650 |
| Device stack | 1.605 | 1.135 | 0.624 | 0.577 | 0.418 |
| Device stack, same absorber | 1.625 | 1.103 | 0.582 | 0.543 | 0.378 |

| Inputs permuted | Random | Level 1 | Level 2 | Level 3 | Temporal |
|---|---|---|---|---|---|
| Fabrication process | 0.958 | 0.665 | 0.455 | 0.450 | 0.371 |
| *Gradient-boosted baseline* | | | | | |
| Composition | 0.629 | 0.905 | 0.373 | 0.318 | 0.433 |
| Device stack | 1.342 | 0.982 | 0.588 | 0.459 | 0.323 |
| Device stack, same absorber | 1.312 | 0.946 | 0.556 | 0.431 | 0.288 |
| Fabrication process | 0.763 | 0.572 | 0.420 | 0.391 | 0.266 |
| *V-PaSS / baseline* | | | | | |
| Composition | 1.49 | 1.04 | 1.57 | 1.34 | 1.50 |
| Device stack | 1.20 | 1.16 | 1.06 | 1.26 | 1.29 |
| Fabrication process | 1.25 | 1.16 | 1.08 | 1.15 | 1.39 |
| *Within V-PaSS* | | | | | |
| Composition / device stack | 0.58 | 0.83 | 0.94 | 0.74 | 1.56 |
| Fabrication process / device stack | 0.60 | 0.59 | 0.73 | 0.78 | 0.89 |
| Composition / fabrication process | 0.98 | 1.41 | 1.29 | 0.95 | 1.75 |
| *Within the baseline* | | | | | |
| Composition / device stack | 0.47 | 0.92 | 0.63 | 0.69 | 1.34 |
| Fabrication process / device stack | 0.57 | 0.58 | 0.71 | 0.85 | 0.83 |
| Composition / fabrication process | 0.82 | 1.58 | 0.89 | 0.81 | 1.63 |

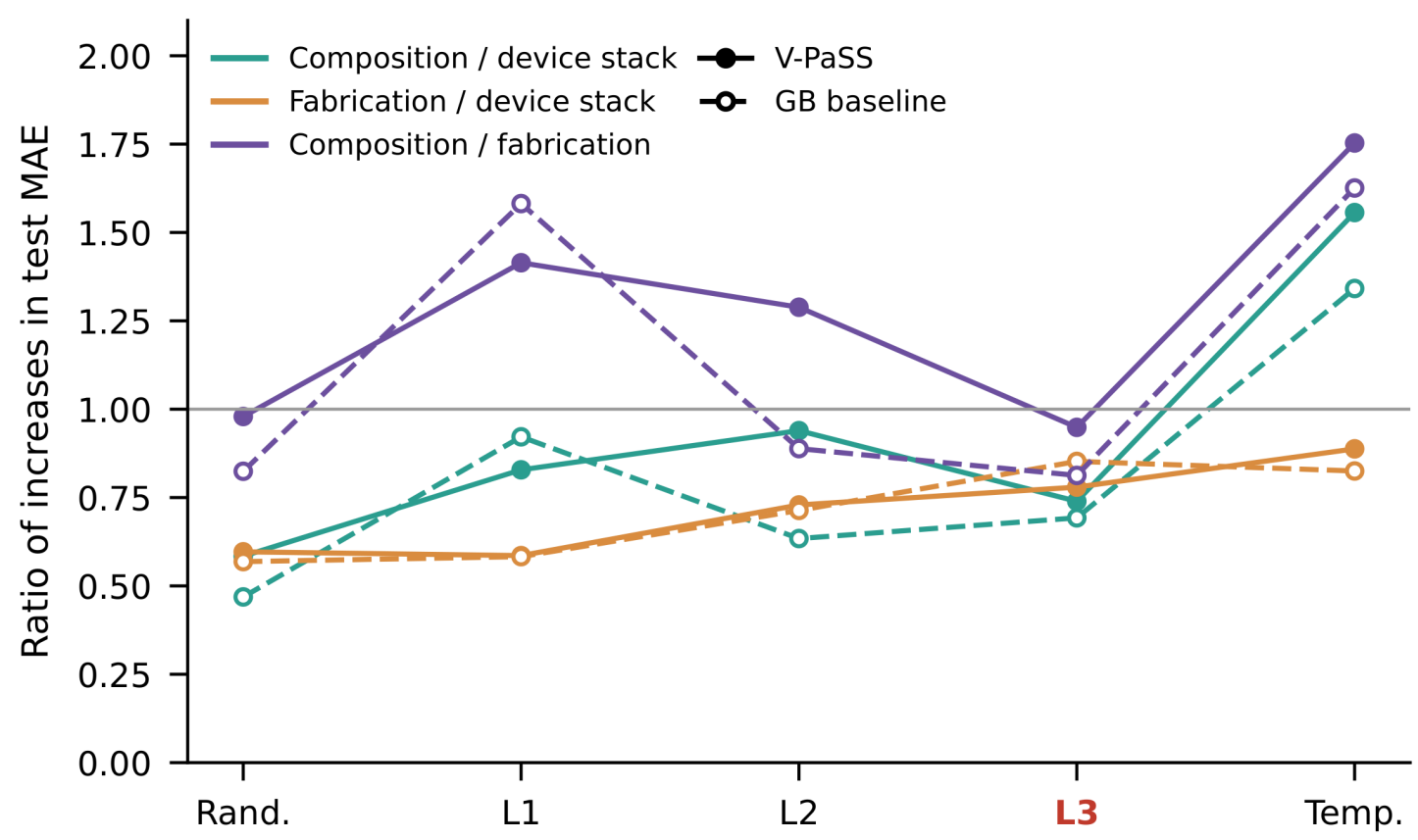

Figure S1: Ratios between the permutation importances of two groups of inputs, within V-PaSS (solid lines) and within the GB baseline (dashed lines), on every split: composition over device stack, fabrication process over device stack, and composition over fabrication process. Data: Table S5.

Table S6: Share (%) of scored test devices whose input is unseen in training, from level 1 to the temporal split, and its growth over that range (temporal share divided by level-1 share). New compositions grow most, and almost all of them are new mixtures of endpoint compounds that occur in training. Composition is the absorber alloy; the device stack is the ETL, HTL, back contact, substrate and polarity together; fabrication is the deposition and additive descriptions together. Layers and fabrication are matched as exact text, so every recorded variant of a layer or a process counts as a different input. On the random split, which shares exact duplicates with training, the shares are 1.9% (composition), 1.1% (fabrication) and 11.6% (device stack).

| Unseen in training | Level 1 | Level 2 | Level 3 | Temporal | Growth, level 1 to temporal |
|---|---|---|---|---|---|
| Composition | 1.1 | 4.5 | 4.7 | 9.5 | 8.6× |
| new mixture of known compounds | 1.1 | 4.4 | 4.6 | 8.8 | 8.1× |
| Fabrication process | 2.1 | 4.7 | 5.0 | 9.1 | 4.2× |
| Device stack | 24.9 | 41.1 | 44.3 | 54.4 | 2.2× |

Table S7: Random split, test MAE (%) on duplicated and new records, and V-PaSS minus each tree-based model on each subset (negative favors V-PaSS). In this split, 1,345 of the 2,362 scored test records (56.9%, the duplicate share reported in the previous study [12]) have an exact duplicate in training and 1,017 do not.

| **Model** | **Duplicated** | **New** | **V-PaSS − model, duplicated** | **V-PaSS − model, new** |
|---|---|---|---|---|
| V-PaSS | 1.886 | 2.244 | — | — |
| LightGBM | 1.792 | 2.213 | +0.094 ± 0.028 | +0.031 ± 0.044 |
| XGBoost | 1.863 | 2.244 | +0.024 ± 0.026 | −0.000 ± 0.044 |
| Random forest | 1.808 | 2.355 | +0.078 ± 0.030 | −0.111 ± 0.048 |
| CatBoost (native) | 2.019 | 2.372 | −0.132 ± 0.030 | −0.128 ± 0.049 |
| Gradient-boosted baseline | 1.970 | 2.344 | −0.084 ± 0.027 | −0.100 ± 0.046 |

Table S8: Temporal split: test MAE (%) on devices that repeat a pre-2016 design and on new designs, and V-PaSS minus each model on each subset (negative favors V-PaSS). Of the 10,752 devices of the temporal split, published from 2018 onward, 3,989 (37.1%) repeat the stack of a device in the pre-2016 training data (**a**), and 1,623 (15.1%) repeat its full recipe, the stack together with the deposition procedure, annealing temperature and additives (**b**).

| **Model** | **Repeated** | **New** | **V-PaSS − model, repeated** | **V-PaSS − model, new** |
|---|---|---|---|---|
| **a**, Repeat of the device stack | | | | |
| V-PaSS | 3.133 | 3.296 | — | — |
| MLP | 3.143 | 3.221 | −0.009 ± 0.061 | +0.075 ± 0.061 |
| TabM | 3.151 | 3.302 | −0.018 ± 0.054 | −0.006 ± 0.061 |
| SciBERT+attention | 3.207 | 3.241 | −0.074 ± 0.050 | +0.055 ± 0.058 |
| XGBoost | 3.254 | 3.598 | −0.120 ± 0.055 | −0.303 ± 0.060 |
| LightGBM | 3.274 | 3.595 | −0.141 ± 0.057 | −0.299 ± 0.060 |
| CatBoost (native) | 3.159 | 3.606 | −0.025 ± 0.050 | −0.310 ± 0.063 |
| Gradient-boosted baseline | 3.258 | 3.582 | −0.124 ± 0.053 | −0.286 ± 0.061 |
| **b**, Repeat of the full recipe | | | | |
| V-PaSS | 3.001 | 3.277 | — | — |
| MLP | 3.020 | 3.223 | −0.018 ± 0.047 | +0.055 ± 0.053 |
| TabM | 3.012 | 3.287 | −0.011 ± 0.047 | −0.010 ± 0.051 |
| SciBERT+attention | 3.084 | 3.254 | −0.083 ± 0.059 | +0.024 ± 0.048 |
| XGBoost | 3.074 | 3.541 | −0.073 ± 0.065 | −0.264 ± 0.051 |
| LightGBM | 3.094 | 3.544 | −0.093 ± 0.069 | −0.267 ± 0.051 |
| Gradient-boosted baseline | 3.033 | 3.538 | −0.032 ± 0.058 | −0.261 ± 0.050 |

Table S9: V-PaSS minus the GB baseline and minus XGBoost (MAE, %; negative favors V-PaSS) on devices with an unseen ETL or HTL and on the rest, and the difference between the two. The XGBoost comparison was not computed on the random split.

| **Split** | **Unseen / rest (devices)** | **Comparator** | **Unseen layer** | **Rest** | **Difference** |
|---|---|---|---|---|---|
| Random | 161 / 2,201 | GB baseline | −0.275 ± 0.138 | −0.077 ± 0.025 | −0.197 ± 0.138 |
| Level 1 | 364 / 2,011 | GB baseline | +0.023 ± 0.173 | +0.050 ± 0.041 | −0.027 ± 0.178 |
| | | XGBoost | +0.098 ± 0.164 | +0.108 ± 0.038 | −0.011 ± 0.168 |
| Level 2 | 1,314 / 3,419 | GB baseline | +0.001 ± 0.120 | +0.009 ± 0.050 | −0.008 ± 0.127 |
| | | XGBoost | +0.004 ± 0.122 | +0.059 ± 0.046 | −0.055 ± 0.128 |
| Level 3 | 1,403 / 3,409 | GB baseline | −0.319 ± 0.109 | −0.089 ± 0.052 | −0.230 ± 0.106 |
| | | XGBoost | −0.213 ± 0.095 | −0.038 ± 0.047 | −0.175 ± 0.095 |
| Temporal | 3,946 / 6,806 | GB baseline | −0.269 ± 0.085 | −0.201 ± 0.047 | −0.068 ± 0.094 |
| | | XGBoost | −0.246 ± 0.083 | −0.229 ± 0.049 | −0.017 ± 0.094 |

Table S10: Test devices with an unseen ETL or HTL: the share with a device from the same publication or institution in training, and V-PaSS minus the GB baseline (MAE, %; negative favors V-PaSS) on those with and without a related training device (same publication at level 1, same institution at level 2), with device counts.

| Split | Unseen-layer devices | Same publication | Same institution | With related device | Without |
|---|---|---|---|---|---|
| Level 1 | 364 | 79.1% | 91.8% | +0.095 ± 0.195 (288) | −0.250 ± 0.383 (76) |
| Level 2 | 1,314 | 0.0% | 89.7% | +0.034 ± 0.123 (1,179) | −0.295 ± 0.419 (135) |
| Level 3 | 1,403 | 0.0% | 0.0% | — | −0.319 ± 0.109 (1,403) |

Table S11: Removing the language-model reading of the device description: change in test MAE (%) against the full V-PaSS with the same three seeds (positive means the reading helped), overall, on mixed absorbers, on devices whose transport-layer name is unseen in training and on the remaining devices, and on devices unseen in training in the stated sense as the difference from the remaining devices or from devices whose transport layers occur in training. A known material under a new name is an unseen transport-layer name whose materials all occur in training; a new transport material is an unseen name containing a material absent from training. New-material devices are also split by whether their publication has a device in training, which is never the case at level 3. The values of Fig. 3d of the main text are the rows for the remaining devices, known materials under a new name and new transport materials.

| Test devices | Level 1 | Level 3 |
|---|---|---|
| All | +0.117 ± 0.046 | +0.151 ± 0.032 |
| Mixed absorbers | +0.149 ± 0.202 | +0.290 ± 0.107 |
| Transport-layer name unseen in training | +0.635 ± 0.227 | +0.344 ± 0.078 |
| Remaining devices | +0.023 ± 0.034 | +0.072 ± 0.038 |
| Transport-layer name unseen in training, minus the rest | +0.612 ± 0.229 | +0.272 ± 0.094 |
| Unseen device stack, minus the rest | +0.387 ± 0.169 | +0.284 ± 0.074 |
| Known materials under a new name (236 / 892 devices) | +0.370 ± 0.220 | +0.353 ± 0.091 |
| Known materials under a new name, minus layers seen in training | +0.347 ± 0.222 | +0.282 ± 0.103 |
| New transport material (128 / 511 devices) | +1.124 ± 0.451 | +0.326 ± 0.150 |
| New transport material, minus layers seen in training | +1.101 ± 0.451 | +0.255 ± 0.160 |
| New transport material, publication with a device in training (102 / 0 devices) | +1.358 ± 0.522 | — |
| New transport material, publication without a device in training (26 / 511 devices) | +0.207 ± 0.673 | +0.326 ± 0.150 |

Table S12: Mixed absorbers. **a**, V-PaSS minus each tree-based model (MAE, %; negative favors V-PaSS). **b**, The reduction in error of V-PaSS relative to the comparator's MAE on the same devices. The number of mixed-absorber test devices is given with each split.

| Split (mixed-absorber devices) | GB baseline | XGBoost | LightGBM | CatBoost (native) |
|---|---|---|---|---|
| **a**, V-PaSS minus the model (MAE, %) | | | | |
| Random (307) | −0.175 ± 0.073 | −0.130 ± 0.067 | −0.110 ± 0.063 | −0.180 ± 0.067 |
| Level 1 (328) | +0.053 ± 0.113 | +0.131 ± 0.089 | +0.102 ± 0.090 | −0.087 ± 0.102 |
| Level 2 (572) | −0.114 ± 0.184 | −0.048 ± 0.199 | −0.084 ± 0.173 | −0.064 ± 0.225 |
| Level 3 (463) | −0.470 ± 0.184 | −0.325 ± 0.180 | −0.399 ± 0.181 | −0.255 ± 0.163 |
| Temporal (1,792) | −0.575 ± 0.112 | −0.633 ± 0.137 | −0.642 ± 0.136 | −0.855 ± 0.142 |
| **b**, reduction in error relative to the model's MAE | | | | |
| Random (307) | 9.7% | 7.4% | 6.4% | 10.0% |
| Level 1 (328) | −2.9% | −7.6% | −5.8% | 4.5% |
| Level 2 (572) | 4.1% | 1.8% | 3.1% | 2.4% |
| Level 3 (463) | 16.1% | 11.7% | 14.0% | 9.4% |
| Temporal (1,792) | 15.8% | 17.1% | 17.3% | 21.8% |

Table S13: Absorber alloys unseen in training, pooled over the five composition-grouped folds, with no measured bandgap for any model: test MAE (%), V-PaSS minus each model (MAE, %; negative favors V-PaSS), and the reduction in error of V-PaSS relative to each model's MAE. The last row is V-PaSS retrained on the same folds with each alloy replaced by its dominant endpoint.

| **Model** | **MAE (2,725)** | **V-PaSS − model (2,725)** | **Relative reduction** | **V-PaSS − model, publication also unseen (1,739)** |
|---|---|---|---|---|
| V-PaSS | 2.587 | — | — | — |
| Gradient-boosted baseline | 3.193 | −0.606 ± 0.106 | 19.0% | −0.695 ± 0.148 |
| XGBoost | 3.089 | −0.502 ± 0.106 | 16.3% | −0.575 ± 0.156 |
| LightGBM | 3.121 | −0.535 ± 0.099 | 17.1% | −0.585 ± 0.140 |
| CatBoost (native) | 3.069 | −0.482 ± 0.101 | 15.7% | −0.509 ± 0.145 |
| V-PaSS, main compound only | 2.792 | −0.205 ± 0.137 | 7.4% | −0.192 ± 0.202 |

Table S14: $CsPbBr_2I$ (92 devices, mean measured efficiency 6.16%, with 56 devices (61%) from a single early study): mean predicted PCE and MAE (%). The tree-based models are the GB baseline, XGBoost, LightGBM and CatBoost with native categorical handling. V-PaSS predicts 11.27% (MAE 5.15%, outperforming all tree models at 7.40–7.80%), while for the iodide-rich counterpart $CsPbI_2Br$, replacement by the dominant endpoint raises the error by 0.762%. Excluding the single inverted-halide device [32] leaves results essentially unchanged (measured mean 6.08%, V-PaSS MAE 5.20%, tree models 7.46–7.88%, and overall alloy construction gain 7.3%).

| **Model** | **Mean predicted PCE** | **MAE** |
|---|---|---|
| Tree-based models | 13.49–13.96 | 7.40–7.80 |
| V-PaSS | 11.27 | 5.15 |
| V-PaSS, main compound only ($CsPbBr_3$) | 7.68 | 2.07 |

Table S15: The reported three-seed GB baseline (MAE, %) compared with the 4,060 three-seed baselines that 30 seeds allow, and the share of these baselines that the reported baseline and V-PaSS are more accurate than.

| **Split** | **Reported** | **Mean of alternatives** | **Range of alternatives** | **Alternatives the reported baseline beats** | **Alternatives V-PaSS beats** |
|---|---|---|---|---|---|
| Level 1 | 2.457 | 2.471 | 2.437–2.519 | 89% | 1% |
| Level 2 | 2.869 | 2.873 | 2.844–2.897 | 65% | 38% |
| Level 3 | 3.017 | 3.008 | 2.988–3.030 | 12% | 100% |

# Methods

**Data and evaluation.** The 23,620 curated devices of the Perovskite Database [7], the publication and institution of each, the five splits, the GB baseline, the twelve tuned comparison models and the bootstrap are those of the previous study [12] and are used unchanged. V-PaSS is trained, validated and scored on the same devices.

**Endpoint library.** The endpoint library holds 23 pure $ABX_3$ compounds (Table S1). Their pristine 3D geometries are from the Materials Project [6]. Devices with more than 15% of their composition mass on compounds absent from the endpoint library are not in the curated set. Of the 23,620 curated devices, 88.5% have an absorber made of a single endpoint compound, and $MAPbI_3$ carries 87.6% of their composition mass. Endpoint band edges are taken from one consistent set of photoemission measurements [17], with ultraviolet photoemission (UPS) for the valence band and inverse photoemission (IPES) for the conduction band. Measurements are used rather than DFT, which systematically underestimates halide perovskite gaps, so that all band offsets share one reference scale. Endpoints not measured there take the values of their nearest analogue in the same source. The two endpoints with no analogue use PBE-DFT band edges from the Materials Project [6] with a single global shift to the vacuum level. The band edges cover 99.99% of the archive's composition mass, and they agree with the median measured bandgap of pure-composition devices in the archive to 0.026 eV overall and 0.012 eV for lead compounds. Goldschmidt tolerance factors $t=(r_A+r_X)/[\sqrt{2}(r_B+r_X)]$ [22] and octahedral factors $\mu = r_B/r_X$ are computed from ionic radii in Å. For MA and FA, the effective radii of Kieslich et al. [24] are used, 2.17 and 2.53. For Cs, Rb, K, Pb, Sn, Ge, I, Br and Cl, Shannon's radii [23] are used (1.67, 1.52, 1.38, 1.19, 1.10, 0.73, 2.20, 1.96 and 1.81 Å). Factors missing for $CsBaI_3$ and $MACuI_3$ are taken from the compound with the most similar formula, $CsGeI_3$ and $MAGeI_3$ respectively.

**Alloy representation.** Each experimental stoichiometry is projected onto the 23 endpoint compounds (Table S1) as described in Section 2 of the main text. The graph network of Eq. 1 of the main text uses multi-head GATv2 layers with interatomic distances as edge features, and an auxiliary head regresses each endpoint's band edges, tolerance factor and octahedral factor from the pooled embedding $p_i$ of Eq. 2 of the main text. The tolerance and octahedral factors are mixed in the same way as the representation and band edges of Eq. 3 of the main text, $t=\sum_i w_i t_i$ and $\mu=\sum_i w_i \mu_i$. The weighted set $\{w_i p_i\}$ is also kept without pooling and serves as keys and values for attention, so each compound in the absorber can be attended to separately.

**Output heads.** In the output heads of Eq. 10 of the main text, $J(E_g)=88.3-45.4E_g+4.35E_g^2$ mA cm$^{-2}$ is a quadratic fit to the Shockley–Queisser photocurrent under the AM1.5G spectrum [26], and the bounds are evaluated at the recorded bandgap, or at the archive's default of 1.60 eV for a device without a recorded bandgap. The four heads are independent predictors. A soft consistency term relates the product of the three factor heads to the PCE head during training, and the PCE head gives the reported efficiency.

**Device stage.** The names of transport layers, additives and substrates are read with a frozen scientific language model (SciBERT [18]), encoded once and cached. The token states pass through a trainable transformer layer and are pooled into one embedding $e$ per field. Additives and dopants are distinguished by an allow-list. The band offsets of Eq. 5 of the main text are assigned by the device polarity, $(\Delta E_{bot}, \Delta E_{top}) = (\Delta E_C, \Delta E_V)$ for n-i-p and $(\Delta E_V, \Delta E_C)$ for p-i-n. The absorber enters the stack of Eq. 6 of the main text through the two tokens of Eqs. 7 and 8 of the main text. The descriptor vector $\varphi$ of Eq. 7 holds the band edges $E_c$ and $E_v$, the bandgap $E_g$, the transport-layer levels $E_{LUMO}{}^{ETL}$ and $E_{HOMO}{}^{HTL}$, the polarity, a flag marking a measured bandgap, $\Omega$, a learned linear map of the $A$-site fractions, the offsets $\Delta E_C$ and $\Delta E_V$, the halide mixing entropy $S_X$, the $A$-site size mismatch $\delta_A$ and the difference between the recorded and the learned bandgap. The $A$-site radii in $\delta_A$ are those of the endpoint library, with 2.78 Å for guanidinium [24] and 2.00 Å for other $A$-site cations. Each token is layer-normalized, and the encoder attends in both directions. The tabular vector $x_{tab}$ of Eq. 9 of the main text holds the bandgap, tolerance and octahedral factors, band offsets, layer thicknesses, process variables and missingness flags, and the substrate and back-contact embeddings enter through low-dimensional projections.

**Training.** Models are trained with AdamW on a cosine learning-rate schedule with warm restarts, with dropout in the fusion heads. Supervision is multi-task (Eq. 11 of the main text): the ratios of $V_{oc}$, $J_{sc}$ and FF to their limits and the standardized PCE, under a Smooth L1 loss. The three ratio terms and the consistency term are combined by learned homoscedastic uncertainty weighting, the PCE term enters with unit weight, and the bandgap term has a fixed weight. Auxiliary supervision on the learned bandgap uses measured values on the 27.7% of records (6,540 of 23,620) that carry one. PCE is standardized as $\widetilde{PCE}=(PCE-\mu_P)/s_P$, with $\mu_P$ and $s_P$ the mean and standard deviation of PCE over the training devices of each split, and the PCE head's output is converted back as $PCE = \mu_P + s_P\, y_P$. Every reported V-PaSS prediction averages five training seeds on the random split (as does the level-1 permutation analysis; Table S5) and the first three seeds on the remaining splits. V-PaSS uses one fixed configuration, which was not tuned and is used unchanged on every split. Each of the twelve tuned benchmark models was instead selected from 30 (random forest, FT-Transformer, TabM and the two SciBERT models), 40 (CatBoost, histogram gradient boosting, MLP and ResNet-MLP) or 60 (XGBoost and LightGBM) configurations by a tree-structured Parzen estimator on level-1 validation data [12].

**Combined models.** Four combinations of V-PaSS with tree-based models were specified before any was scored. Three are weighted averages of V-PaSS with one tree-based model, with the weight chosen on a grid. Their partners are the GB baseline, XGBoost, and XGBoost given V-PaSS's input representation (the endpoint weights and the physical inputs computed without the network). The fourth is a non-negative combination of V-PaSS and the six tuned tree-based models summing to one, fitted by least absolute error. All weights are fitted on validation data, the tree models are refitted with their stored configurations and reproduce the test predictions of the previous study, and the test set is scored once. Every combination is reported on every split (Table S4).

**Permutation importance and component ablations.** Reliance of V-PaSS and of the GB baseline on each group of inputs is measured by permutation importance [31]. One group at a time is reassigned at random among the test devices while the other inputs stay in place. This is repeated five times per seed, with the seeds of each published model. The groups follow the feature-group analysis of the previous study [12]. The composition comprises the endpoint weights, site fractions, formula, band edges, recorded bandgap and structural factors. The device stack comprises the names of the substrate, transport layers and back contact and their energy levels and thicknesses. For V-PaSS, it also includes the stack polarity, which assigns the role of each layer that V-PaSS reads by position. The fabrication process comprises the deposition and additive descriptions and the annealing, anti-solvent and solvent-annealing treatments. To separate the device stack from its correlation with the composition, the stack is also permuted exclusively among test devices sharing the same absorber composition (affecting 89–92% of scored devices). Because most test devices share the same absorber, mainly $MAPbI_3$, a permutation leaves the inputs of many devices unchanged, so each importance is scored only on the devices whose inputs the permutation changes. A composition counts as changed when the alloy or the recorded bandgap of the donor device differs, and a stack when any of its layers or its polarity differs. Permutation supplies contradictory rather than absent inputs, so its magnitudes are compared between models and splits. Component ablations retrain V-PaSS with one component removed, using the configuration, splits and first three seeds of the full V-PaSS, and compare it with the full V-PaSS on the same seeds and test devices, ensuring that devices differing only in the ablated feature receive identical inputs. The language-model ablation replaces every device text field (substrate, transport layers, back contact, additives, deposition) with its own fixed random token sequence for each distinct string at levels 1 and 3, preserving categorical identity, tabulated energy levels and parsed numerical features while stripping semantic meaning.

**Test-device subsets.** Mixed absorbers are those whose composition has non-zero weight on more than one endpoint compound. On these devices, V-PaSS is compared with the GB baseline and with the three tuned tree-based models that have the lowest error at level 3, namely XGBoost, LightGBM and CatBoost with native categorical handling (Table S12). To assess generalization beyond familiar designs, the test devices of each split are partitioned into repeated and new subsets, and the performance margin of V-PaSS over comparator models is evaluated on each. On the random split, a test record is duplicated when it has an exact duplicate in training (Table S7). On the temporal split, a test device repeats a design when it repeats the stack of a device in the pre-2016 training data (the same absorber composition, ETL, HTL, back contact, substrate and architecture), and repeats the full recipe when it also has the same deposition procedure, anneal temperature and additives (Table S8). A test device has an unseen transport layer when its ETL or HTL never occurs among the training records of that split (Table S9). At levels 1 to 3, such a device has a related training device when a device from the same publication or the same institution is in training (Table S10). In the language-model ablation, the devices with a transport-layer name unseen in training are divided further. In some, the transport layer contains a material unseen in training. In the others, the name recombines materials that occur in training, for example a known oxide with a new additive or an extra sub-layer (Table S11).

**Test on alloys unseen in training.** In this test, all devices with the same mixed composition are kept together, so no test alloy occurs in training or validation. Every pure composition stays in training as one of the building blocks from which alloys are constructed. The 567 distinct mixed compositions are counted as distinct sets of endpoint fractions. They are shuffled with a fixed seed and packed by device count into five folds of 545 devices. For fold $k$, the devices of fold $k$ form the test set and those of fold $k$+1 the validation set. The training set holds every pure-composition device and the remaining three folds. Each fold serves once as the test set, so that every one of the 2,725 mixed-absorber devices (including the 1,739 devices whose publication is likewise absent from training) is predicted by a model that has never seen its alloy. No model receives the recorded bandgap in this test: V-PaSS treats every gap as unmeasured, and the tree-based models drop the bandgap column. V-PaSS uses its first seed in each fold, and the tree-based models average three seeds of their published configurations, with target encodings computed from each fold's training devices. In the alloy-construction ablation, each alloy is replaced by its dominant endpoint, with the same folds and seed, and its cost is reported over devices, over alloys and per fold.

**Uncertainty estimation.** Uncertainties are given as ± one standard error. The standard error is the width of the 95% joint bootstrap interval (20,000 resamples) divided by 3.92, and the bootstrap resamples each split's grouping unit [12], never single devices. In the test on alloys unseen in training, the bootstrap resamples the test alloys. Training-seed variation is reported separately and never pooled with test uncertainty. The GB baseline has a fixed configuration and was retrained with 30 seeds at each of levels 1 to 3. The first three seeds reproduce the reported predictions exactly, and the 30 seeds give 4,060 three-seed baselines (Table S15). Unless stated otherwise, ± in tables and figures denotes the standard error defined here.

**Data availability.** The data supporting the findings are in the paper and its Supplement. The splits, the provenance of every device and the predictions of the thirteen benchmark models are released with the previous study [12].

# References

[1] Tian Xie and Jeffrey C. Grossman. Crystal graph convolutional neural networks for an accurate and interpretable prediction of material properties. *Physical Review Letters*, 120(14):145301, 2018.

[2] Rhys E A Goodall and Alpha A Lee. Predicting materials properties without crystal structure: Deep representation learning from stoichiometry. *Nature Communications*, 11:6280, 2020.

[3] Anthony Yu-Tung Wang, Steven K Kauwe, Ryan J Murdock, and Taylor D Sparks. Compositionally restricted attention-based network for materials property predictions. *npj Computational Materials*, 7:77, 2021.

[4] Simon Batzner, Albert Musaelian, Lixin Sun, et al. E3-equivariant graph neural networks for data-efficient and accurate interatomic potentials. *Nature Communications*, 13(1):2453, 2022.

[5] Vahe Tshitoyan, John Dagdelen, Leigh Weston, Alexander Dunn, Ziqin Rong, Olga Kononova, Kristin A Persson, Gerbrand Ceder, and Anubhav Jain. Unsupervised word

embeddings capture latent knowledge from materials science literature. *Nature*, 571(7763):95–98, 2019.

[6] Anubhav Jain, Shyue Ping Ong, Geoffroy Hautier, Wei Chen, William Davidson Richards, Stephen Dacek, Shreyas Cholia, Dan Gunter, David Skinner, Gerbrand Ceder, et al. Commentary: The materials project: A materials genome approach to accelerating materials innovation. *APL materials*, 1(1):011002, 2013.

[7] T. Jesper Jacobsson, Adam Hultqvist, Alberto García-Fernández, et al. An open-access database and analysis tool for perovskite solar cells based on the FAIR data principles. *Nature Energy*, 7(1):107–115, 2022.

[8] Shuxia Tao, Shanshan Cao, Xinxin Tang, et al. Machine learning for perovskite materials design and discovery. *npj Computational Materials*, 7(1):23, 2021.

[9] X. Wang et al. Graph neural network-assisted machine learning for high-throughput buried interfacial materials screening in antisolvent-free perovskite solar cells. *ACS Applied Materials & Interfaces*, 2026.

[10] Wenhao Yang, Jiangang Li, Xin Chen, Yanling Feng, Chenhui Wu, Ilangovan Gunasekaran, Chao Fan, and Wenqiang Li. Data-driven design of high-performance $MASn_xPb_{1-x}I_3$ perovskite materials by machine learning and experimental realization. *Light: Science & Applications*, 11:232, 2022.

[11] Anagha Aneesh, Nawaf Alampara, José Márquez, and Kevin Maik Jablonka. Semantic device graphs for perovskite solar cell design. In AI for Accelerated Materials Design Workshop (AI4MAT), ICLR, 2025. OpenReview AGCClISEXL.

[12] Qiwei Han and Chi Zhou. Provenance-grouped protocol for evaluating machine learning in perovskite photovoltaics. 2026. arXiv:2610.12766.

[13] Edward J. Beard and Jacqueline M. Cole. Perovskite- and dye-sensitized solar-cell device databases auto-generated using ChemDataExtractor. *Scientific Data*, 9:329, 2022.

[14] Sherjeel Shabih, Hampus Näsström, Sharat Patil, Asmin Askin, Keely Dodd-Clements, Jessica Helisa Hautrive Rossato, Hugo Gajardoni de Lemos, Yuxin Liu, Florian Mathies, Natalia Maticiuc, Rico Meitzner, Edgar Nandayapa, Juan José Patiño López, Yaru Wang, Lauri Himanen, Eva Unger, T. Jesper Jacobsson, José A. Márquez, and Kevin Maik Jablonka. An autonomous living database for perovskite photovoltaics. *arXiv preprint arXiv:2601.17807*, 2026.

[15] L Bellaiche and David Vanderbilt. Virtual crystal approximation revisited: Application to dielectric and piezoelectric properties of perovskites. *Physical Review B*, 61(12):7877, 2000.

[16] Mudasser Ahmad, Ghulam Rehman, Liaqat Ali, Muhammad Shafiq, Rashid Iqbal, Riaz Ahmad, Tahir Khan, S Jalali-Asadabadi, M Maqbool, and Iftikhar Ahmad. Influence of halide composition on the structural, electronic, and optical properties of mixed $CH_3NH_3Pb(I_{1-x}Br_x)_3$ perovskites calculated using the virtual crystal approximation method. *Physical Review B*, 94:125139, 2016.

[17] Shuxia Tao, Ines Schmidt, Geert Brocks, Junke Jiang, Ionut Tranca, Klaus Meerholz, and Selina Olthof. Absolute energy level positions in tin- and lead-based halide perovskites. *Nature Communications*, 10(1):2560, 2019.

[18] Iz Beltagy, Kyle Lo, and Arman Cohan. SciBERT: A pretrained language model for scientific text. *arXiv preprint arXiv:1903.10676*, 2019.

[19] Shaked Brody, Uri Alon, and Eran Yahav. How attentive are graph attention networks? In International Conference on Learning Representations (ICLR), 2022.

[20] Simon Kahmann, Zehua Chen, Oleh Hordiichuk, Olga Nazarenko, Shuyan Shao, Maksym V. Kovalenko, Graeme R. Blake, Shuxia Tao, and Maria A. Loi. Compositional variation in $FAPb_{1-x}Sn_xI_3$ and its impact on the electronic structure: A combined density functional theory and experimental study. *ACS Applied Materials & Interfaces*, 14(30):34253–34261, 2022.

[21] Kimberley J. Savill, Aleksander M. Ulatowski, and Laura M. Herz. Optoelectronic properties of tin–lead halide perovskites. *ACS Energy Letters*, 6(7):2413–2426, 2021.

[22] V. M. Goldschmidt. Die Gesetze der Krystallochemie. *Die Naturwissenschaften*, 14(21):477–485, 1926.

[23] R. D. Shannon. Revised effective ionic radii and systematic studies of interatomic distances in halides and chalcogenides. *Acta Crystallographica Section A*, 32(5):751–767, 1976.

[24] Gregor Kieslich, Shijing Sun, and Anthony K. Cheetham. Solid-state principles applied to organic–inorganic perovskites: New tricks for an old dog. *Chemical Science*, 5(12):4712–4715, 2014.

[25] William Shockley and Hans J. Queisser. Detailed balance limit of efficiency of p-n junction solar cells. *Journal of Applied Physics*, 32(3):510–519, 1961.

[26] International Electrotechnical Commission. Photovoltaic devices — part 3: Measurement principles for terrestrial photovoltaic (PV) solar devices with reference spectral irradiance data. IEC 60904-3:2019, 2019.

[27] Sven Rühle. Tabulated values of the Shockley–Queisser limit for single junction solar cells. *Solar Energy*, 130:139–147, 2016.

[28] Uwe Rau, Beatrix Blank, Thomas C. M. Müller, and Thomas Kirchartz. Efficiency potential of photovoltaic materials and devices unveiled by detailed-balance analysis. *Physical Review Applied*, 7:044016, 2017.

[29] Martin A. Green. Solar cell fill factors: General graph and empirical expressions. *Solid-State Electronics*, 24:788–789, 1981.

[30] Alex Kendall, Yarin Gal, and Roberto Cipolla. Multi-task learning using uncertainty to weigh losses for scene geometry and semantics. In 2018 IEEE/CVF Conference on Computer Vision and Pattern Recognition, pages 7482–7491, 2018.

[31] Leo Breiman. Random forests. *Machine Learning*, 45(1):5–32, 2001.

[32] Hongrui Sun, Jing Zhang, Xinlei Gan, Luting Yu, Haobo Yuan, Minghui Shang, Chaojie Lu, Dagang Hou, Ziyang Hu, Yuejin Zhu, and Liyuan Han. Pb-reduced $CsPb_{0.9}Zn_{0.1}I_2Br$ thin films for efficient perovskite solar cells. *Advanced Energy Materials*, 9(25):1900896, 2019.